\documentclass[lettersize,journal]{IEEEtran}
\usepackage{amsmath,amsfonts, amssymb}
\usepackage{algorithm, algpseudocode}
\usepackage{array}
\usepackage[caption=false,font=normalsize,labelfont=sf,textfont=sf]{subfig}
\usepackage{textcomp}
\usepackage{stfloats}
\usepackage{url}
\usepackage{verbatim}

\usepackage{graphicx}
\usepackage{hyperref}
\usepackage{cite}
\usepackage{booktabs}
\usepackage{mathtools}
\usepackage{placeins}
\usepackage{xcolor}
\newtheorem{theorem}{Theorem}
\DeclareMathOperator*{\argmin}{arg\,min}

\begin{document}

\title{Multivariate Fields of Experts for Convergent Image Reconstruction}

\author{Stanislas Ducotterd and Michael Unser}



\maketitle
\renewcommand*{\thefootnote}{\fnsymbol{footnote}}
\footnotetext[1]{The authors are with the Biomedical Imaging Group, \'Ecole polytechnique f\'ed\'erale de Lausanne (EPFL), {\text \{forename.name\}@epfl.ch}. }
\renewcommand*{\thefootnote}{\arabic{footnote}}

\begin{abstract}
We introduce the multivariate fields of experts, a new framework for the learning of image priors. Our model generalizes existing fields of experts methods by incorporating multivariate potential functions constructed via Moreau envelopes of the $\ell_\infty$-norm. We demonstrate the effectiveness of our proposal across a range of inverse problems that include image denoising, deblurring, compressed-sensing magnetic-resonance imaging, and computed tomography. The proposed approach outperforms comparable univariate models and achieves performance close to that of deep-learning-based  regularizers while being significantly faster, requiring fewer parameters, and being trained on substantially fewer data. In addition, our model retains a high level of interpretability due to its structured design. It is supported by theoretical convergence guarantees which ensure reliability in sensitive reconstruction tasks.
\end{abstract}


\section{Introduction}

A common problem in science and engineering is to recover an object of interest from indirect linear measurements. Let $\mathbf H \in \mathbb{R}^{m \times n}$ denote the measurement operator and let $\mathbf y \in \mathbb{R}^m$ be the observed data. The goal is to reconstruct the underlying signal $\mathbf x \in \mathbb{R}^n$ such that $\mathbf H \mathbf x \approx \mathbf y$. However, since $\mathbf{y}$ is usually noisy and $\mathbf{H}$ is often ill-conditioned or even rank-deficient, the direct inversion is typically unstable and results in poor reconstructions. To address this, a popular strategy is to use variational regularization to recover $\mathbf x$ as the minimizer of the energy

\begin{equation}\label{eq:main_obj}
f(\mathbf x) = \tfrac{1}{2}\|\mathbf H \mathbf x - \mathbf y\|_2^2 + \lambda R(\mathbf x),
\end{equation}
where the first term enforces data fidelity and the regularizer $R \colon \mathbb R^n \to \mathbb R$ encodes prior knowledge about the underlying signal.

Many popular regularizers \cite{roth_fields_2005, 6705653, 8119908, 10223264, 23M1565243, zach_product_2024, Pourya03072024} can be expressed as a sum of univariate potential functions applied to filter responses, a structure frequently referred to as the fields of experts (FoE) model. In particular, the weakly convex ridge regularizer (WCRR) \cite{23M1565243} represents a recent advance in this direction, where nonconvex potentials are learned via linear splines. While highly effective, such approaches implicitly assume independence between the filter responses (channels), ignoring potentially valuable interactions between them.

In this work, we propose the multivariate fields of experts (MFoE), a novel class of learned regularizers that generalizes the classic FoE framework by incorporating multivariate potential functions that have the capability to capture interactions between channels. Our contributions are as follows.

\begin{itemize}
\item \textbf{Multivariate Generalization:} We extend the WCRR framework to the multivariate setting by introducing a novel class of parametric potentials based on Moreau envelopes.
\item \textbf{Tailored Optimization:} We design a specialized optimization algorithm for the resulting objective and prove its convergence to a stationary point.
\item \textbf{Comprehensive Validation:} We demonstrate the effectiveness of our approach across diverse inverse problems that include denoising, deblurring, compressed-sensing magnetic-resonance imaging (CS-MRI), and computed tomography (CT). We demonstrate superior performance to univariate FoE methods. Our code is accessible on Github \footnote{\href{https://github.com/StanislasDucotterd/MFoE}{https://github.com/StanislasDucotterd/MFoE}}.
\end{itemize}

\section{Related Work}

A popular model for image reconstruction is the total variation (TV) regularizer \cite{rudin_nonlinear_1992}, where the regularization term $R_{\text{TV}}$ promotes piecewise-constant solutions by penalizing the horizontal and vertical finite differences of the image. Since $R_{\text{TV}}$ is not differentiable, the objective \eqref{eq:main_obj} is typically minimized via the proximal operator
\begin{equation}\label{eq:obj_tv}
    \operatorname{prox}_{\lambda R}(\mathbf y) = \argmin_{\mathbf x \in \mathbb R^n} \left(\tfrac{1}{2}\|\mathbf x - \mathbf y\|_2^2 + \lambda R_{\text{TV}}(\mathbf x)\right),
\end{equation}
which yields the solution of a denoising problem—the simplest instance of \eqref{eq:main_obj} with $\mathbf H = \mathbf I$. This proximal operator does not admit a closed-form solution and must be computed through iterative algorithms \cite{tv_chambolle}. 

To further improve the reconstruction performance of variational methods, many works propose data-driven approaches to construct $R$. The FoE framework \cite{roth_fields_2005} relies on a regularizer defined for any image $\mathbf x \in \mathbb R^n$ as
\begin{equation}\label{eq:ridge_reg}
    R_{\text{FoE}}(\mathbf x) = \sum_{k=1}^K \langle \mathbf 1_n, \psi_k(\mathbf W_k \mathbf x)\rangle,
\end{equation}
where $\mathbf W_k \in \mathbb R^{n \times n}$ denotes a convolution matrix and $\psi_k\colon \mathbb R \to \mathbb R$ is a univariate nonlinearity. In the original FoE model, the potentials $\psi_k$ correspond to the negative log-likelihood of Student-$t$ distributions and the filters $\{\mathbf{W}_k\}_{k=1}^K$ are learned by the minimization of $\sum_{m=1}^M R_{\text{FoE}}(\mathbf x_m)$ over a dataset $\{\mathbf x_m\}_{m=1}^M$ of natural images.

A more advanced strategy for the learning of the regularizer is proposed in \cite{6705653}, where bilevel optimization \cite{stackelberg} is employed. In this setting, the filters are trained explicitly to minimize the denoising loss
\begin{equation}\label{eq:denoising_loss}
    \sum_{m=1}^M \|\operatorname{prox}_{\lambda R_{\text{FoE}}}(\mathbf x_m + \mathbf n_m) - \mathbf x_m\|_2^2,
\end{equation}
where $\mathbf n_m$ is an additive white Gaussian noise. This method aligns the training objective with the evaluation metric (PSNR), which yields better results than the likelihood-based learning of the original FoE.

Although the potentials $\{\psi_k\}_{k=1}^K$ are generally nonconvex in FoE, which makes the proximal operator $\operatorname{prox}_{\lambda R}$ potentially set-valued, bilevel optimization remains applicable and effective in practice.

Subsequent works \cite{8119908, 10223264, 23M1565243, zach_product_2024, Pourya03072024} increase the expressivity of the FoE model further by jointly learning the filters and the nonlinearities $\{\psi_k\}_{k=1}^K$. The WCRR learns the gradient of the regularizer by parameterizing the derivatives of univariate potentials with linear splines. Our model relies on higher-dimensional potential functions to extend this approach to the multivariate setting.

While the approaches discussed above minimize an explicit energy functional, a parallel line of research known as Plug-and-Play (PnP) \cite{6737048} bypasses the use of an explicit regularizer. Instead, PnP methods insert an off-the-shelf denoiser as a surrogate for the proximal operator in iterative optimization algorithms. Although PnP yields good performance, it generally does not correspond to the minimization of an explicit objective and often lacks convergence guarantees. Several strategies have been proposed to address this issue, such as the enforcement of Lipschitz constraints on the denoiser \cite{ryu2019plug, ducotterd_jmlr} or the use of continuation schemes \cite{7744574}.

Finally, a recent line of work has bridged the gap between variational methods and the PnP framework by parameterizing an explicit regularizer with deep neural networks \cite{hurault_gs, hurault_proximal_2022, fang_whats_2024}. These methods learn the regularizer via its gradient, which requires the computation of second-order derivatives during training. This yields highly expressive regularizers with an explicit objective and general convergence guarantees, but comes at the cost of a significant computational overhead—both during training, due to the need to differentiate the objective twice, and at inference time, since the gradient of the neural network must be evaluated at each optimization step.

\section{Description of the Model}

\subsection{Notations and Definitions}

The least-squares projection onto a set $\mathcal{C}$ is defined as 

\begin{equation}
    \operatorname{Proj}_{\mathcal{C}}(\mathbf x) = \argmin_{\mathbf z \in \mathcal{C}} \|\mathbf z - \mathbf x\|_2^2.
\end{equation}
This projection exists and is unique whenever the set $\mathcal{C}$ is closed and convex. A particular closed convex set of interest in this work is the $\ell_1$-ball in $\mathbb{R}^d$, which we denote by
\begin{equation}
    \mathcal{B}_{\ell_1} = \left\{ \mathbf{z} \in \mathbb{R}^d : \|\mathbf{z}\|_1 \leq 1 \right\}.
\end{equation}

\noindent
The Moreau envelope \cite{moreau1965proximite} of a proper, lower semicontinuous, convex function $f \colon \mathbb{R}^d \to \mathbb{R}$ with parameter $\mu > 0$ is defined as
\begin{equation} \label{eq:moreau}
    M_{f}^{\mu}(\mathbf{x}) = \inf_{\mathbf{z} \in \mathbb{R}^d} \left( f(\mathbf{z}) + \frac{1}{2\mu} \|\mathbf{z} - \mathbf{x}\|_2^2 \right).
\end{equation}
The gradient of the Moreau envelope of $f$ is related to the proximal operator of $f$ through the equality

\begin{equation}\label{eq:moreau_prox}
    \nabla M_{f}^\mu(\mathbf x)=\frac{1}{\mu}\left(\mathbf x-\operatorname{prox}_{\mu f}(\mathbf x)\right).
\end{equation}
In this paper, we denote the Moreau envelope (with parameter $\mu$) of the $\ell_\infty$-norm in $\mathbb{R}^d$ as $\rho^d_\mu \colon \mathbb{R}^d \to \mathbb{R}_{\geq 0}$. We rely on the standard identity \cite[Equation 6.8]{parikh2014proximal}

\begin{equation}\label{eq:prox_proj}
\operatorname{prox}_{\mu\|\cdot\|_\infty}(\mathbf x) = \mathbf x - \mu\operatorname{Proj}_{\mathcal{B}_{\ell_1}}(\mathbf x / \mu).
\end{equation}
The proximal operator explicitly realizes the infimum in the definition of the Moreau envelope. Therefore, the evaluation of~\eqref{eq:moreau} at $\mathbf{z} = \operatorname{prox}_{\mu\|\cdot\|_\infty}(\mathbf x)$ yields

\begin{equation}\label{eq:potential}
    \rho_\mu^d(\mathbf x) = \|\mathbf x - \mu\operatorname{Proj}_{\mathcal{B}_{\ell_1}}(\mathbf x / \mu)\|_{\infty} + \frac{\mu}{2}\|\operatorname{Proj}_{\mathcal{B}_{\ell_1}}(\mathbf x / \mu)\|_2^2.
\end{equation}
From~\eqref{eq:moreau_prox}, we then obtain the convenient formula for the gradient
\begin{equation}\label{eq:grad_potential}
    \nabla \rho_\mu^d(\mathbf{x}) = \operatorname{Proj}_{\mathcal{B}_{\ell_1}}(\mathbf{x}/\mu),
\end{equation}
which can be computed efficiently via sorting using the Condat algorithm \cite{condat2016fast}.

\subsection{Generalization of the WCRR}\label{sec:gen_wcrr}

\begin{figure}
\centering
\includegraphics[scale=0.28]{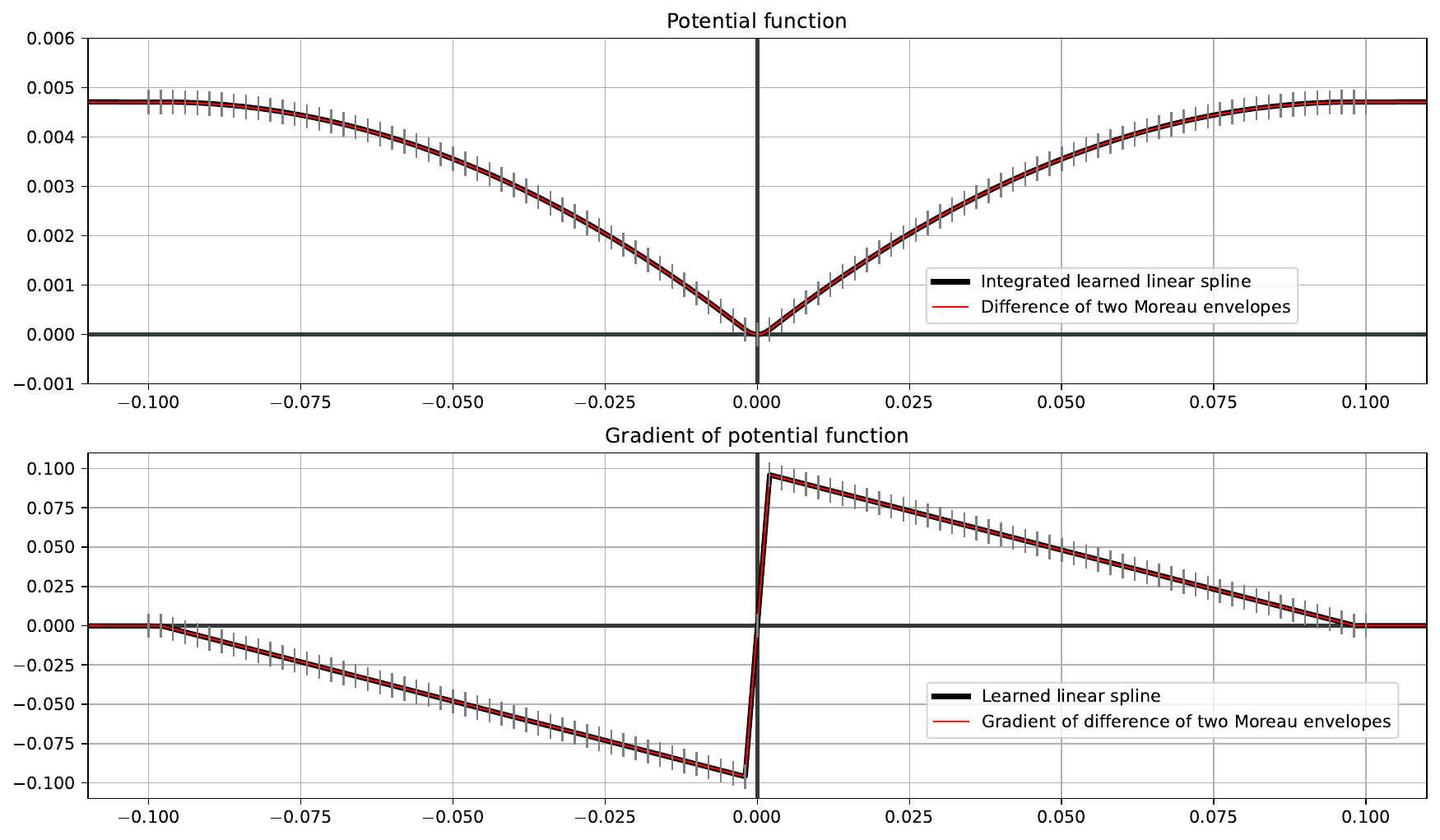}
\caption{The WCRR potential (top) and its gradient (bottom) match the difference of two Moreau envelopes of the absolute value. Vertical ticks indicate the spline knots.}
\label{fig:comparison}
\end{figure}

The WCRR \cite{23M1565243} constructs a regularization functional by applying 1-weakly convex, univariate potential functions parameterized as linear splines to learned filter responses. Our proposed model generalizes this framework by replacing these scalar functions with higher-dimensional analogs, a generalization directly motivated by the geometric properties of the WCRR potentials. Specifically, these potentials take the form $\psi_k(x) = \alpha_k^{-2}\psi(\alpha_k x)$ where a single base function $\psi$ is shared across all channels. In Figure~\ref{fig:comparison}, we illustrate that $\psi$ (which is learned via a linear spline with more than 100 knots) can be approximated almost to perfection by a simple three-parameter function of the form $t \mapsto \lambda(\rho_\mu^1(t) - \rho_\nu^1(t))$ with $\mu = 0.002$ and $\lambda = \nu = 0.098$. This observation motivates the use of $\rho^d_\mu$ as a foundation for a multivariate extension, with the property that the model can reduce to the WCRR when $d = 1$. This three-parameter function is closely related to the minimax-concave penalty introduced in \cite{7938377}.

Note that the absolute value is the only one-dimensional function (up to scaling) that is a norm. This suggests to extend it by the consideration of the Moreau envelopes of general norms. We focus on norms for which both the Moreau envelope and its gradient can be computed efficiently. The $\ell_1$-norm decomposes into a sum of absolute values, which implies that the difference of two Moreau envelopes recovers the univariate formulation of the WCRR. 

The $\ell_2$ and $\ell_\infty$ norms are particularly tractable. We select the $\ell_\infty$-norm for two reasons: first, it yields better empirical results; second, the authors of \cite[Theorem 4]{unser2025} establish that any norm on $\mathbb{R}^d$ can be approximated to arbitrary precision by the function $\mathbf{x} \mapsto \|\mathbf{L}\mathbf{x}\|_\infty$, where $\mathbf{L}$ maps into a potentially higher-dimensional space. This universality property ensures that our model can represent a broad class of convex potentials simply by learning the appropriate linear filters.

The WCRR constrains its potentials to be 1-weakly convex to ensure overall convexity during denoising, given the 1-strong convexity of the data term. Our optimization algorithm
allows us to relax this constraint in our model without losing any convergence guarantees.

\subsection{General Setting}\label{sec:model}

We now move beyond the univariate setting and extend the FoE framework to the multivariate case, where the regularizer is defined for any image $\mathbf x \in \mathbb R^n$ as

\begin{equation}\label{eq:mfoe}
    R(\mathbf{x}) = \sum_{k=1}^{K}\left\langle\mathbf{1}_n, \psi_{k}^d\left(\mathbf{W}_{k}^d \mathbf{x}\right)\right\rangle,
\end{equation}
where $\mathbf W_k^d \in \mathbb R^{(d n) \times n}$ denotes a multi-convolution matrix that maps a single-channel image to a $d$-channel one, and $\psi_k^d\colon \mathbb R^d \to \mathbb R$ is a nonlinearity given by

\begin{equation}\label{eq:weak_potential}
    \psi_{k}^d(\mathbf x) = \mu_k\rho_{\mu_k}^d(\mathbf x) - \mu_k\rho_{\tau_k\mu_k}^d(\mathbf Q_k \mathbf x),
\end{equation}
where $\mathbf Q_k \in \mathbb R^{d \times d}$ and $\tau_k$ is a scalar. \\

\begin{theorem}\label{thm:single_global_min}
    \textit{If $\mathbf Q_k$ is such that $\|\mathbf Q_k\|_{\infty} \leq 1$ and $\tau_k > \|\mathbf Q_k\|_2^2$, then the nonlinearity~\eqref{eq:weak_potential} is nonnegative with a unique global minimum at the origin, and a gradient that is nonexpansive.}
\end{theorem}

Theorem \ref{thm:single_global_min} guarantees that the regularizer is bounded from below and preserves a interpretation similar to the classic FoE, where weak filter responses are favored while stronger responses are penalized. The proof can be found in Appendix~\ref{sec:proof}.

The gradient of our regularizer can be written as 

\begin{equation}\label{eq:gradient}
    \mathbf \nabla R(\mathbf x) = \mathbf W^T\varphi(\mathbf{Wx}),
\end{equation}
where $\mathbf W \in \mathbb R^{(K d n) \times n}$ refers to the concatenation of all the $d$-channel convolutions and $\varphi$ refers to the concatenation of all the gradients $\mathbf \nabla \psi_k^d\colon \mathbb R^d \to \mathbb R^d$. The spectral norm $\|\mathbf W\|_2$ of the filters is enforced to be one which ensures that $\mathbf \nabla R$ is nonexpansive.

To efficiently train filters with a large receptive field, we decompose $\mathbf{W}$ into a three zero-padded convolutions with small kernels and an increasing number of output channels, as in \cite{10223264}. Similarly to \cite{9156693}, the filters are constrained to have zero mean.

\subsection{Optimization}

We design an algorithm that minimizes the objective function defined in \eqref{eq:main_obj} and guarantees convergence to a stationary point. We build upon the work of \cite{ipiano} which uses the heavy-ball method \cite{POLYAK19641}
\begin{equation}
    \mathbf x_{k+1} = \mathbf x_k - \alpha_k \nabla f(\mathbf x_k) + \beta_k(\mathbf x_k - \mathbf x_{k-1}).
\end{equation}
The method of \cite{ipiano} ensures convergence by imposing strict constraints on the stepsize $\alpha_k$ and momentum $\beta_k$. Here, we relax these constraints to allow for more aggressive acceleration while still maintaining convergence guarantees, with the backing of Theorem~\ref{thm:convergence} below. As described in Algorithm~\ref{alg:hbr} (Lines 5--6), our refinement is a backtracking mechanism that discards the inertial steps and falls back to a standard gradient descent update if the candidate update $\mathbf x_{k+1}$ fails to satisfy a sufficient descent condition.

The backtracking step introduces negligible overhead because the cost of the check of the descent condition is minimal. This is due to the specific structure of our potentials
\begin{equation}
    \rho^d_\mu(\mathbf x) = \|\mathbf x - \mu\nabla \rho^d_\mu(\mathbf{x})\|_{\infty} + \frac{\mu}{2}\|\nabla \rho^d_\mu(\mathbf{x})\|_2^2.
\end{equation}
Since $R$ is composed of these potentials, the function value $R(\mathbf x)$ can be computed cheaply alongside the gradient $\nabla R(\mathbf x)$. Furthermore, the values computed during a rejected candidate step are reused for the fallback step, which ensures that the objective and gradient are evaluated twice only when a restart occurs.

The nonexpansiveness of $\mathbf \nabla R$ implies that $\mathbf \nabla f$ is $L$-Lipschitz continuous with constant
\begin{equation}
    L = \|\mathbf H\|_2^2 + \lambda.
\end{equation}

\begin{theorem}\label{thm:convergence}
    \textit{Let $\alpha \in (0, 2/L)$. If the sequence $\{\mathbf x_k\}_{k \geq 0}$ generated by Algorithm \ref{alg:hbr} is bounded, then it converges to a stationary point of $f$ and has a finite length, with
    \begin{equation}
        \sum_{k\geq 0} \|\mathbf x_{k+1} - \mathbf x_k\|_2 < \infty.
    \end{equation}}
\end{theorem}

Theorem~\ref{thm:convergence} is essential to the reliability of image reconstruction, as the finite-length property guarantees that the iterates settle to a specific solution rather than being indefinitely oscillating or diverging. The proof can be found in Appendix~\ref{sec:proof2}. The theorem assumes bounded iterates, a condition that can also be satisfied formally by augmenting the objective with a coercive barrier term such as $B(\mathbf{x}) = \log\left(1 + \exp\left(\|\mathbf{x}\|_2^2 - C\right)\right)$. There, the constant term $C$ can be chosen arbitrarily large. The coercive term ensures bounded level sets while it remains negligible within the valid range of image intensities. Similar ideas have been proposed in \cite{23M1565243} and \cite{hurault_proximal_2022}.

Throughout this work, we use the fixed stepsize $\alpha = 1.99/L$. We increase the momentum parameter $\beta_k$ after a successful inertial step and reset it to a base value otherwise, a common heuristic in accelerated optimization.

\begin{algorithm}[t]
\caption{Heavy-Ball with Restart}
\label{alg:hbr}
\begin{algorithmic}[1] 
\Statex \hspace*{-\algorithmicindent} \textbf{Input:} Function $f$ with $L$-Lipschitz gradients, stepsize $\alpha$, initialization $\mathbf{x}_0 \in \mathbb{R}^n$, tolerance $\varepsilon > 0$
\State Set $k = 0$, $\beta_0 = 0.5$, $r = \infty$
\While{$r > \varepsilon$}
    \State $\mathbf{x}_{k+1} = \mathbf{x}_{k} - \alpha \mathbf \nabla f(\mathbf{x}_{k}) + \beta_k\left(\mathbf x_k - \mathbf x_{k-1}\right)$
    \State $\beta_{k+1} = \tfrac{1}{2}(\beta_k + 1)$
    \If{$f(\mathbf x_{k+1}) > f(\mathbf x_k) - \alpha\left(1 - \frac{L\alpha}{2}\right)\|\nabla f(\mathbf x_k)\|_2^2$}
        \State $\mathbf{x}_{k+1} = \mathbf{x}_k - \alpha \mathbf \nabla f(\mathbf x_k)$
        \State $\beta_{k+1} = 0.5$ 
    \EndIf
    \State $r = \|\mathbf{x}_{k+1} - \mathbf{x}_k\| / \|\mathbf{x}_{k+1}\|$
    \State $k \gets k + 1$
\EndWhile
\Statex \hspace*{-\algorithmicindent} \textbf{Output:} Approximate solution $\mathbf{x}_{k+1}$
\end{algorithmic}
\end{algorithm}

\section{Training on Denoising}\label{sec:train_denoising}

We adopt a bilevel optimization strategy to learn the parameters
\begin{equation}
    \boldsymbol \theta = \{\mathbf W, \{\mathbf Q_k, \tau_k\}_{k=1}^K, \boldsymbol \theta_{\mathbf f}, \lambda\},
\end{equation}
where $\boldsymbol \theta_{\mathbf f}$ parametrizes the function $\mathbf f$ from Section~\ref{sec:noise_level}. To enforce the constraints $\|\mathbf W\|_2 = 1$, $\|\mathbf Q_k\|_\infty \leq 1$, and $\tau_k > \|\mathbf Q_k\|_2^2$, we optimize over unconstrained auxiliary variables mapped to the feasible set via
\begin{equation}
\begin{aligned}
    \mathbf W &= \frac{\tilde{\mathbf W}}{\|\tilde{\mathbf W}\|_2}, \quad 
    \mathbf Q_k = \frac{\tilde{\mathbf Q}_k}{\max(1, \|\tilde{\mathbf Q}_k\|_\infty)}, \\
    \tau_k &= 1.001 \exp((\tilde{\tau}_k)_+) \cdot \|\mathbf Q_k\|_2^2,
\end{aligned}
\end{equation}
where the spectral norm $\|\tilde{\mathbf W}\|_2$ is computed via the Fourier transform \cite{23M1565243}.

\subsection{Bilevel Formulation}

To make our regularizer more universal, we train it on a range of Gaussian-noise levels and denote by $R_\sigma$ the regularizer associated with the noise level $\sigma$. We use a dataset of paired samples $(\mathbf x_m, \mathbf y_m)_{m=1}^M$, where $\mathbf x_m$ is the reference image and $\mathbf y_m = \mathbf x_m + \sigma_m \mathbf n_m$ with $\mathbf n_m \sim \mathcal{N}(\mathbf 0, \mathbf I)$ and $\sigma_m \sim \mathcal{U}(0, 0.2)$. This yields the bilevel optimization problem
\begin{align}
    \min_{\boldsymbol \theta} \quad & \mathcal L(\boldsymbol \theta) = \frac{1}{M}\sum_{m=1}^M \frac{1}{\sqrt{\sigma_m}}\Vert \mathbf x_m^*(\boldsymbol \theta) - \mathbf x_m \Vert_1 \label{eq:outer_loss} \\
    \text{s.t.} \quad & \mathbf x_m^*(\boldsymbol \theta) = \argmin_{\mathbf x \in \mathbb R^n} \left( \frac{1}{2}\|\mathbf x - \mathbf y_m\|_2^2 + \lambda R_{\sigma_m}(\mathbf x) \right). \label{eq:inner_prob}
\end{align}
The dependence on $\sigma$ is only enforced through the parameters $\{\mu_k\}_{k=1}^K$, while the filters and other components remain the same across all noise levels. We describe their parameterization in Appendix~\ref{sec:noise_level}.

The weighting factor $1/\sqrt{\sigma_m}$ ensures that the loss remains sensitive to errors at low noise levels. This is a common strategy used in the training of diffusion models \cite{song2021scorebased}. 

\subsection{Inner Optimization (Forward Pass)}

The inner problem consists of the minimization of \eqref{eq:inner_prob}. Since this objective is not necessarily convex, we find an approximate stationary point $\mathbf x_m^*$ by running the heavy-ball with restart (see Algorithm~\ref{alg:hbr}) until the relative change of the iterates falls below $10^{-4}$. Since $\mathbf \nabla f(\mathbf x_m^*) \approx 0$, the point $\mathbf x_m^*$ can be viewed as the fixed point of the gradient-step operator

\begin{equation}\label{eq:fixed_point}
    \mathbf x_m^* \approx \mathbf x_m^* -\alpha \mathbf \nabla f(\mathbf x_m^*) = \mathcal{T}(\mathbf x_m^*).
\end{equation}

\subsection{Outer Optimization (Backward Pass)}

We need access to the gradient $\nabla_{\boldsymbol \theta} \mathcal L$ to minimize the outer objective \eqref{eq:outer_loss} via gradient descent. Applying the chain rule, we have that
\begin{equation}
    \nabla_{\boldsymbol \theta} \mathcal L = \sum_{m=1}^M \left( \frac{\partial \mathbf x_m^*}{\partial \boldsymbol \theta} \right)^T \nabla_{\mathbf x_m^*} \mathcal{L}.
\end{equation}
As shown in the deep equilibrium framework \cite{NEURIPS2019_01386bd6}, one can compute the gradient of $\mathbf x_m^*$ with respect to $\boldsymbol \theta$ using the implicit function theorem applied to \eqref{eq:fixed_point} as
\begin{equation}\label{eq:implicit_diff}
    \frac{\partial \mathbf x_m^*}{\partial \boldsymbol \theta} = \left( \mathbf I - \mathbf J_{\mathcal{T}}(\mathbf x_m^*) \right)^{-1} \frac{\partial \mathcal{T}}{\partial \boldsymbol \theta},
\end{equation}
where $\mathbf J_{\mathcal{T}}$ is the Jacobian of the fixed-point operator with respect to $\mathbf x$.

The determination of the inverse in \eqref{eq:implicit_diff} is computationally prohibitive. Instead, we approximate it using 25 steps of the Broyden algorithm \cite{broyden_class_1965}, as implemented in the \texttt{torchdeq} library \cite{torchdeq}. This allows us to backpropagate through the equilibrium point without unrolling the entire optimization trajectory, which results in significant reduction of the memory usage.

\subsection{Implementation Details}

The training is conducted on a dataset of 238,400 small patches of size $\left(40 \times 40\right)$, extracted from 400 images from the BSD500 dataset \cite{arbelaez_contour_2011} and scaled to $[0,1]$. The number of training images is relatively small compared to typical deep-learning-based methods, which often rely on significantly bigger datasets to achieve good generalization \cite{9454311}. We use the ADAM optimizer with a batch size of 128 and set the learning rate to 0.005 for the filters $\mathbf W$ and the function $\boldsymbol \mu$, while all other parameters use a learning rate of 0.05. The model is trained for 5,000 steps and the learning rates are decayed by a factor of 0.75 every 500 steps. The full training procedure takes about 5 hours on a Tesla V100 GPU.

\subsection{Results and Ablation Studies}

We report the denoising performance of MFoE in Table~\ref{tab:denoising-results} on the grayscale-converted BSD68 \cite{martin2001database}, McMaster \cite{zhang2011color} and Set14 \cite{zeyde2010single} datasets. We use $K = 15$ potentials of input dimension $d = 4$ with filters of size $\left(11 \times 11\right)$. For comparison purposes, we include TV \cite{rudin_nonlinear_1992}, the WCRR \cite{23M1565243}, and Prox-DRUNet \cite{hurault_proximal_2022}. The TV regularizer is a classic baseline with a very simple architecture. Prox-DRUNet is the most advanced deep-learning-based regularizer. It involves three order of magnitudes more parameters than the MFoE. 

Note that the WCRR results differ slightly from those in the original paper because we trained it on noise levels ranging from 0 to $50/255$, whereas the original implementation was ranging from 0 to $30/255$. We achieved this by modifying only the splines $s_{\alpha_i}$ from \cite[Equation~(17)]{23M1565243}, increasing the number of knots from 11 to 18 to maintain interval lengths as close as possible to the original configuration.

As mentioned in Section~\ref{sec:gen_wcrr}, we add the relaxed version of WCRR (WCRR-free) and an alternative version of our model based on the $\ell_2$-norm (MFoE-$\ell_2$). More details about their implementation can be found in Appendices~\ref{sec:wcrr_free} and \ref{sec:mfoe_l2}. 

The relaxation of the 1-weak convexity constraint in WCRR yields clear improvements in performance, as evidenced by WCRR-free consistently outperforming the constrained WCRR baseline. MFoE-$\ell_2$ achieves results comparable to the unconstrained univariate model. However, the proposed MFoE model consistently outperforms all univariate baselines (WCRR and WCRR-free) as well as the MFoE-$\ell_2$ variant, which confirms the benefits of the learned multivariate $\ell_\infty$-based potentials. Notably, all four models employ the same number of filters $\left(K d = 60\right)$ and have roughly the same number of parameters, which ensures that the observed performance gains are not due to increased model capacity but result instead from the greater expressivity of the proposed multivariate potentials. Although Prox-DRUNet maintains a performance lead, the proposed MFoE reduces this margin while offering a principled, interpretable architecture backed by formal convergence guarantees.

We present in Table~\ref{tab:denoising-results2} the performance of the MFoE as we increase the number of filters from $K d = 60$ to 240. Unlike the WCRR, which saturates at 60 filters, our model continues to improve and reaches a performance plateau around 180 filters. To ensure fair comparisons with the WCRR throughout the paper, we adopt the 60-filter variant as our default model.

We further investigate the tradeoff between $K$ and $d$ in Table~\ref{tab:denoising-results3}, keeping the total number of filters fixed at the product $K d = 60$. We observe a significant improvement when moving from $d = 1$ to $d = 2$, with performance peaking at $d = 4$. Beyond that, performance degrades steadily, with $d = 60$ performing worse than even the univariate case $d = 1$. We attribute the drop in performance to the nature of the $l_{\infty}$-norm. The gradient is nonzero only for the filter with the maximal response within each group. Fewer filters get updated at each step as $K$ is decreased. This dynamic can cause filters that rarely achieve the maximal response to stagnate at their initialization.

\begin{table*}[t]
\centering
\caption{Denoising performance (PSNR / SSIM) on the BSD68, McMaster, and Set14 datasets.}
\footnotesize 
\setlength{\tabcolsep}{2.5pt} 
\begin{tabular}{l ccc ccc ccc c}
\toprule
 & \multicolumn{3}{c}{BSD68} & \multicolumn{3}{c}{McMaster} & \multicolumn{3}{c}{Set14} & \\
\cmidrule(lr){2-4} \cmidrule(lr){5-7} \cmidrule(lr){8-10}
Method & $\sigma=15$ & $\sigma=25$ & $\sigma=50$ & $\sigma=15$ & $\sigma=25$ & $\sigma=50$ & $\sigma=15$ & $\sigma=25$ & $\sigma=50$ & Params. \\
\midrule
TV 
    & 29.90 / 0.911 & 27.48 / 0.857 & 24.83 / 0.771 
    & 31.51 / 0.930 & 29.07 / 0.888 & 26.28 / 0.822 
    & 30.19 / 0.915 & 27.71 / 0.864 & 24.86 / 0.782 
    & 1 \\
WCRR 
    & 31.20 / 0.933 & 28.68 / 0.890 & 25.68 / 0.806 
    & 33.24 / 0.952 & 30.61 / 0.920 & 27.23 / 0.849 
    & 31.74 / 0.938 & 29.19 / 0.900 & 25.93 / 0.817 
    & $1.4 \cdot 10^4$ \\
WCRR-free 
    & 31.18 / 0.932 & 28.68 / 0.888 & 25.77 / 0.807 
    & 33.32 / 0.953 & 30.77 / 0.923 & 27.53 / \underline{0.861} 
    & 31.76 / 0.937 & 29.27 / 0.899 & 26.12 / 0.821 
    & $1.4 \cdot 10^4$ \\
MFoE-$\ell_2$
    & 31.21 / 0.933 & 28.71 / 0.889 & 25.80 / 0.809 
    & 33.32 / 0.953 & 30.77 / 0.923 & 27.50 / 0.860 
    & 31.77 / 0.938 & 29.28 / 0.900 & 26.08 / 0.822 
    & $1.4 \cdot 10^4$ \\
MFoE 
    & \underline{31.32} / \underline{0.935} & \underline{28.84} / \underline{0.892} & \underline{25.93} / \underline{0.812} 
    & \underline{33.53} / \underline{0.955} & \underline{31.02} / \underline{0.926} & \underline{27.73} / \textbf{0.865} 
    & \underline{31.96} / \underline{0.940} & \underline{29.54} / \underline{0.904} & \underline{26.35} / \underline{0.827} 
    & $1.4 \cdot 10^4$ \\
Prox-DRUNet 
    & \textbf{31.70} / \textbf{0.940} & \textbf{29.18} / \textbf{0.898} & \textbf{26.14} / \textbf{0.815} 
    & \textbf{33.92} / \textbf{0.957} & \textbf{31.32} / \textbf{0.927} & \textbf{27.97} / 0.860 
    & \textbf{32.50} / \textbf{0.945} & \textbf{30.06} / \textbf{0.911} & \textbf{26.83} / \textbf{0.835} 
    & $1.7 \cdot 10^7$ \\
\bottomrule
\end{tabular}
\label{tab:denoising-results}
\end{table*}

\begin{table}[t]
\centering
\caption{Denoising performance in terms of noise level on the BSD68 test set.}
\setlength{\tabcolsep}{5pt}
\begin{tabular}{lccc}
\toprule
& $\sigma=15/255$ & $\sigma=25/255$ & $\sigma=50/255$ \\
\hline
MFoE ($K=15$, $d=4$) & 31.32 & 28.84 & 25.92 \\
MFoE ($K=30$, $d=4$) & 31.36 & 28.88 & 25.97 \\
MFoE ($K=45$, $d=4$) & 31.38 & \textbf{28.90} & \textbf{25.99} \\
MFoE ($K=60$, $d=4$) & \textbf{31.39} & \textbf{28.90} & 25.97 \\
\bottomrule
\end{tabular}
\label{tab:denoising-results2}
\end{table}

\section{Visualization and Analysis of the Learned Model}

\begin{figure*}
\centering
\includegraphics[scale=0.65]{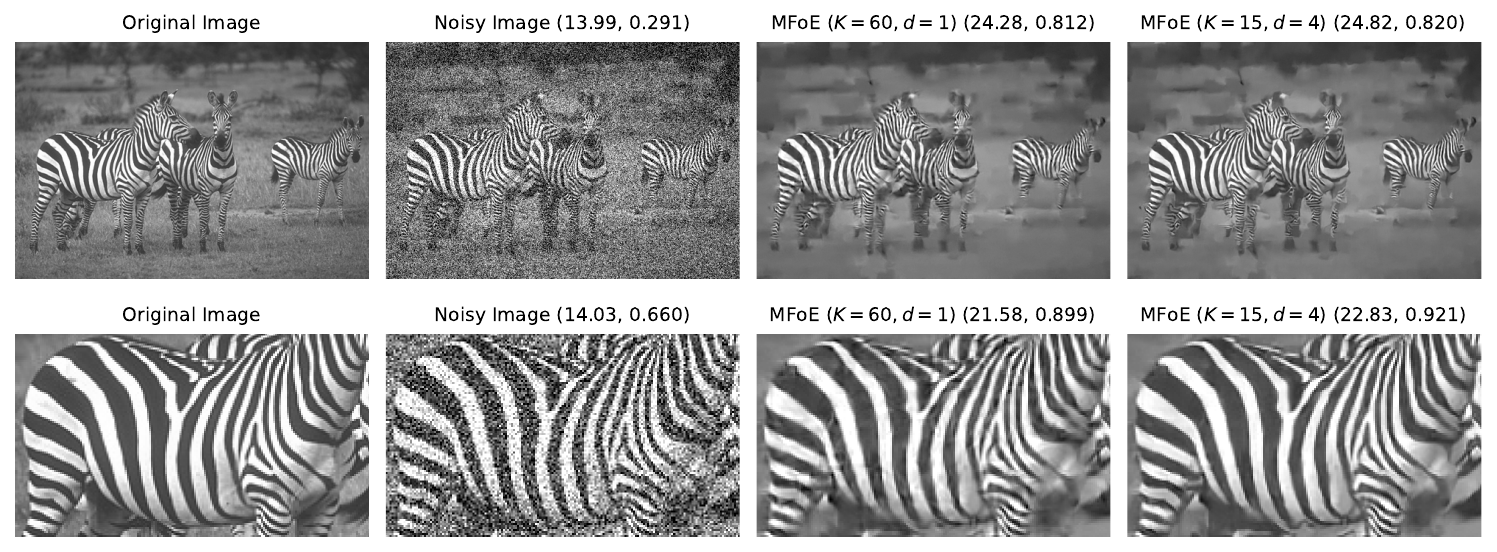}
\caption{Denoising results for \( \sigma = 0.2 \), shown on the full image and on a zoomed-in textured region. The reported metrics are (PSNR, SSIM).}
\label{fig:zebras}
\end{figure*}

A qualitative comparison of our denoising results with the univariate baselines reveals that our approach performs particularly well on periodic patterns, as shown in Figure~\ref{fig:zebras} on the zebra stripes. 


\begin{figure}
\centering
\includegraphics[scale=0.55]{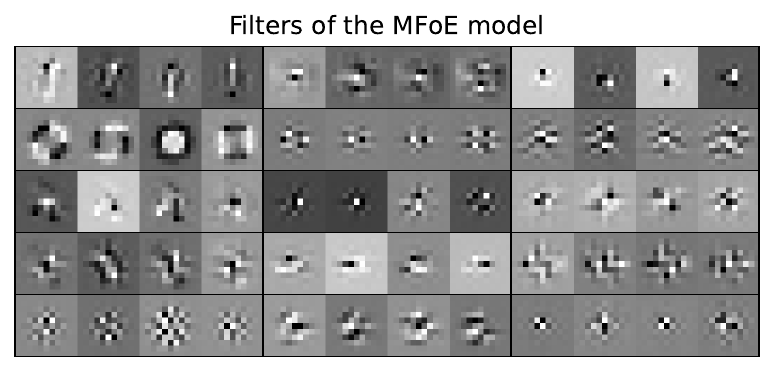}
\caption{Learned filters on the denoising task with $K = 15$ and $d = 4$. The black lines split the filters into the 15 different groups of 4.}
\label{fig:mfoe_filters}
\end{figure}

We attribute this improvement to the mechanics of quadrature filtering. Consider a pure harmonic signal $x(t) = \cos(\omega_0 t)$. A single linear filter $h$ transforms this signal into $(h * x)(t) = A \cos(\omega_0 t + \varphi)$, where $A = |H(\omega_0)|$ and $\varphi = \arg(H(\omega_0))$ denote the spectral magnitude and phase, respectively. Because the response of the filter oscillates, the regularization penalty induced by a univariate potential will fluctuate spatially. This behavior may result in an uneven signal reconstruction. To avoid this effect, one may use a pair of quadrature filters $(h_1, h_2)$ \cite{adelson1985spatiotemporal}. These are built such that $|H_1(\omega_0)| = |H_2(\omega_0)| = A$ and $\arg(H_2(\omega_0)) = \arg(H_1(\omega_0)) + \frac{\pi}{2}$. By combining their output via a square root, we get
\begin{align}
    & \sqrt{(h_1 * x)^2 + (h_2 * x)^2} \nonumber \\
    & \quad = \sqrt{A^2 \cos^2(\omega_0 t + \varphi) + A^2 \sin^2(\omega_0 t + \varphi)} \nonumber \\
    &\quad = A.
    \label{eq:quadrature_energy}
\end{align}
This is a configuration that cannot be achieved by a univariate architecture.

\begin{figure}
\centering
\includegraphics[scale=0.35]{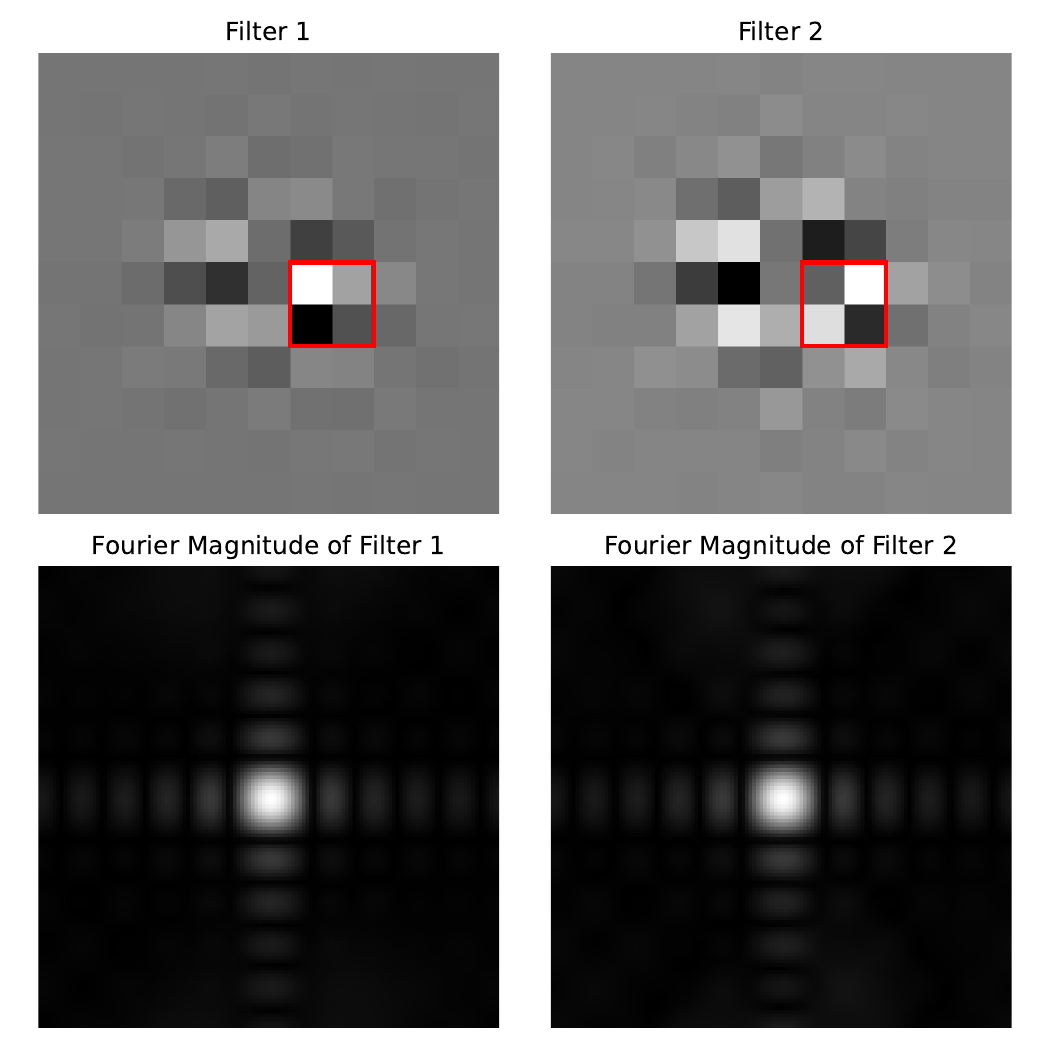}
\caption{A pair of learned filters from the $\left(K = 30 \text{ and } d = 2\right)$ model. In the top row, a red frame highlights a location where the two filters differ most markedly. The magnitude of the Fourier transforms are shown in the bottom row.}
\label{fig:magnitude_filters}
\end{figure}

The filters of the $(K=15, d=4)$ model displayed in Figure~\ref{fig:mfoe_filters} exhibit similar and complementary structures within the same group. Those filters can work together to extract meaningful patterns in schemes that are reminiscent of quadrature-filter pairs and, more generally, steerable filterbanks \cite{freeman1991design, 5357447}. We show a group of two filters and their Fourier magnitudes from the $(K=30, d=2)$ model in Figure~\ref{fig:magnitude_filters}. Those two filters are almost identical except for a horizontal shift of a group of two pixels, as highlighted by the red frame. Because of this, they have very similar Fourier magnitudes, which suggests some quadrature mechanism within this group.


\begin{figure}
\centering
\includegraphics[scale=0.3]{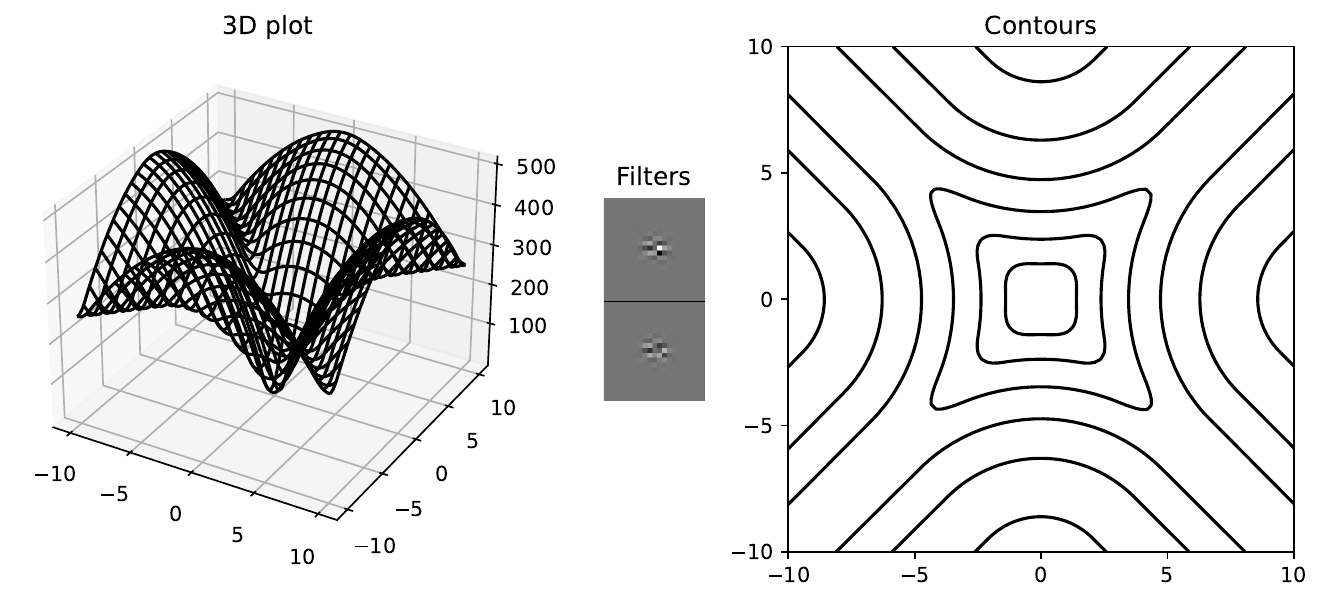}
\caption{Learned potential function and the corresponding filters. The axes correspond to the responses of the two filters.}
\label{fig:potential0}
\end{figure}

We plot the learned potential of those same two filters in Figure~\ref{fig:potential0}. Visually, the learned potential resembles a rotated $\ell_p$-norm with $p < 1$, a geometry characteristic of antisparsity \cite{elvira2017bayesian}. The regularizer penalizes correlated responses less severely than independent ones. This property favors joint activations and avoids the detrimental periodic nature of individual filter responses via a coupling similar to that of quadrature filters.


\begin{figure}
\centering
\includegraphics[scale=0.25]{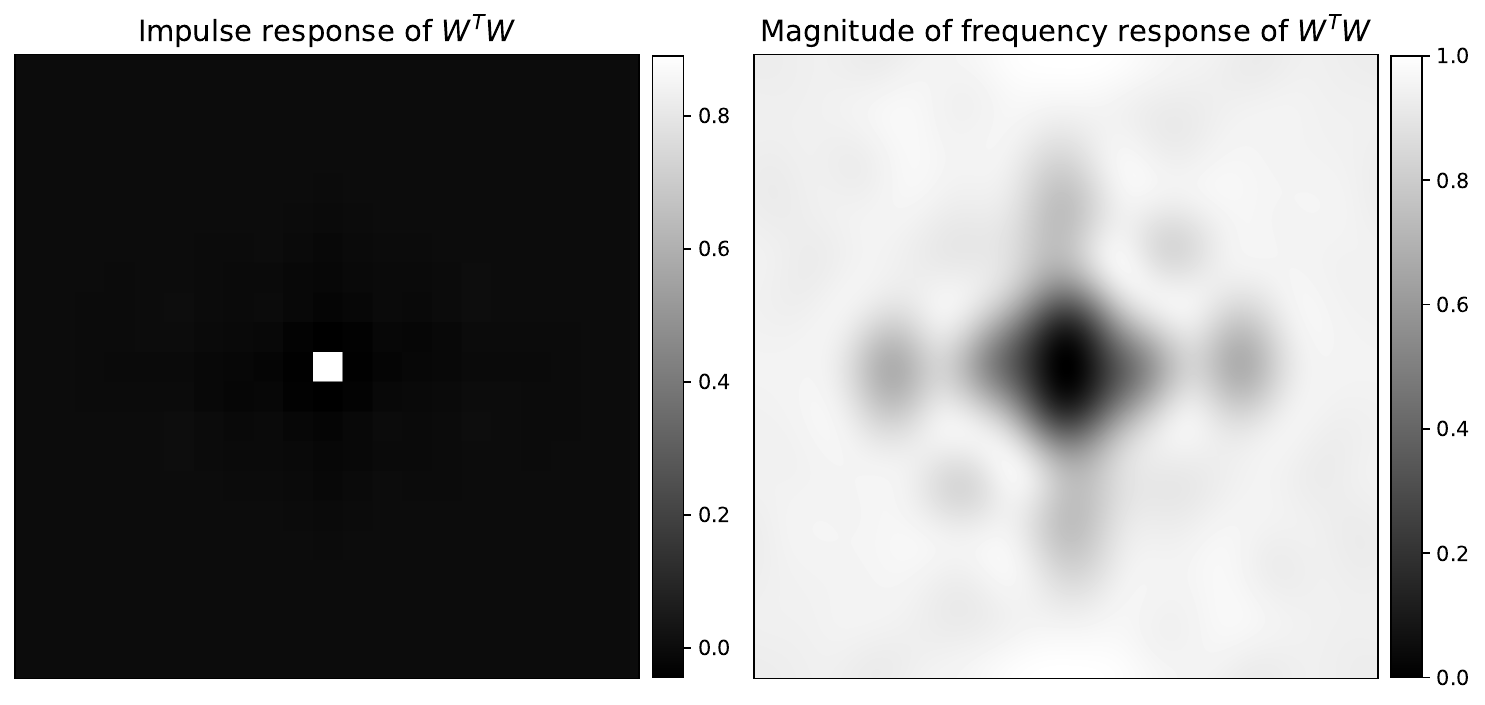}
\caption{Impulse and frequency response of $\mathbf W^T \mathbf W$. The image size on the left is $\left(21 \times 21\right)$, identical to the size of $\mathbf W^T \mathbf W$. The image size on the right is $\left(1500 \times 1500\right)$, which accomodates for some amount of zero-padding in the Fourier transform.}
\label{fig:mfoe_impulse}
\end{figure}

A corollary of Theorem~\ref{thm:single_global_min} is that the minimizers of our regularizer lie in the null space of $\mathbf{W}$. In Figure~\ref{fig:mfoe_impulse}, we show the impulse response of the filter that corresponds to $\mathbf{W}^T \mathbf{W}$, along with the magnitude of its Fourier transform. While the impulse response is close to a Kronecker impulse, it cannot be exactly so as it must sum up to zero due to the zero-mean constraint on all filters in $\mathbf{W}$. The frequency response reveals that only the zero-frequency coefficient vanishes. This suggests that the null space of $\mathbf{W}$ is one-dimensional and consists of constant images. 



\section{Inverse Problems}

We benchmark MFoE for image deblurring, CS-MRI, and CT. For each task, we simulate measurements as
\begin{equation}
    \mathbf y = \mathbf{Hx} + \mathbf w,
\end{equation}
where $\mathbf H$ is the linear operator corresponding to the imaging modality and $\mathbf w \sim \mathcal{N}(\mathbf 0, \sigma_{\mathbf w}^2\mathbf I)$.

We compare our learned regularizer with TV, WCRR, and Prox-DRUNet. To reconstruct the images, we minimize the objective \eqref{eq:main_obj} using the solvers recommended in the respective original works; namely, proximal gradient descent for TV, safeguarded accelerated gradient descent \cite{23M1565243} for WCRR, Douglas-Ratchford splitting \cite{douglas_numerical_1956} for Prox-DRUNet, and the heavy-ball method with restart (Algorithm~\ref{alg:hbr}) for MFoE. In all cases, the algorithm is terminated when the relative difference between consecutive iterates falls below $10^{-5}$ or when the number of iterations exceeds 1,000.

To deploy the models, we tune their hyperparameters on a validation set using the coarse-to-fine grid search from \cite{10223264}.
Then, the performance is reported on a dedicated test set.
For TV, we only tune the regularization strength $\lambda$  in \eqref{eq:main_obj}. For the other methods, we tune $\lambda$ and the noise level $\sigma$. 

\subsection{Deblurring}

We apply the frameworks to a deblurring task, where $\mathbf H$ is a convolution operator. We take three kernels of size (25 $\times$ 25) from \cite{5206815}, as shown on the top of Table~\ref{tab:deblurring}. The first kernel is isotropic Gaussian while the other two are anisotropic and model a motion blur. We consider the noise levels $\sigma_{\mathbf w} \in \{0.01, 0.03\}$. Each method is initialized with the observed blurred and noisy measurements. The hyperparameters are tuned on the Set12 \cite{zhang2017beyond} dataset and the performance is reported in Table~\ref{tab:deblurring} on the BSD68 dataset. 

\begin{figure*}
\centering
\includegraphics[scale=0.85]{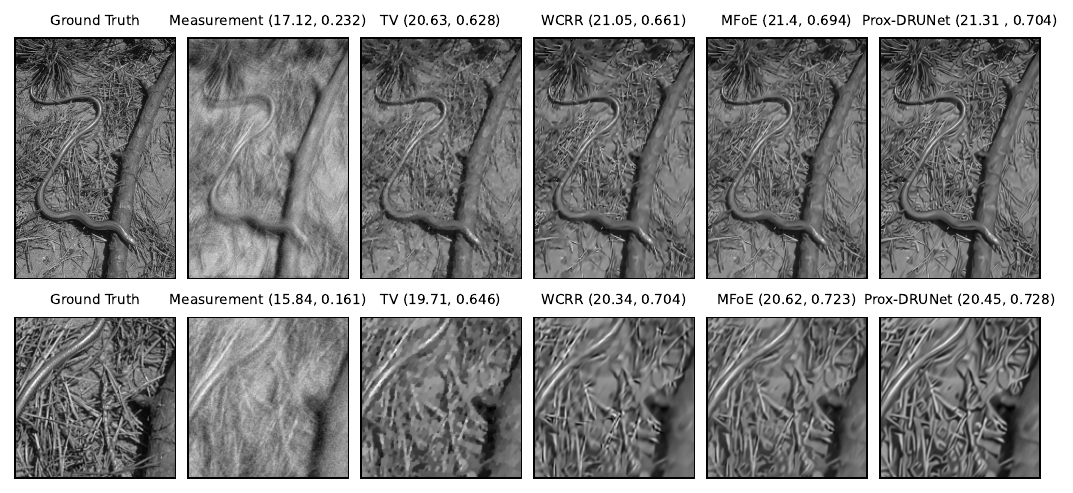}
\caption{Reconstruction for a blur with the third kernel and $\sigma_{\mathbf w} = 0.03$. The reported metrics are (PSNR, SSIM).}
\label{fig:deblurring}
\end{figure*}

The MFoE outperforms the WCRR in every instance and manages to match the performance of Prox-DRUNet for the isotropic blur. Examples of deblurring are provided in Figure~\ref{fig:deblurring}. There, we can see in the zoomed part that Prox-DRUNet makes a somewhat bolder prediction than the other models. This results in a higher SSIM but a lower PSNR than MFoE, despite the blurrier prediction of MFoE.

\begin{table*}[t]
\centering
\caption{Quantitative results (PSNR / SSIM) for deblurring on the BSD68 dataset.}
\label{tab:deblurring}
\setlength{\tabcolsep}{6pt} 
\begin{tabular}{lcccccc}
\toprule
& \multicolumn{2}{c}{\includegraphics[scale=1.0]{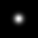}} 
 & \multicolumn{2}{c}{\includegraphics[scale=1.0]{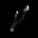}} 
 & \multicolumn{2}{c}{\includegraphics[scale=1.0]{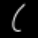}} \\
\cmidrule(lr){2-3} \cmidrule(lr){4-5} \cmidrule(lr){6-7}
 & $\sigma_{\mathbf{w}} = 0.01$ & $\sigma_{\mathbf{w}} = 0.03$ 
 & $\sigma_{\mathbf{w}} = 0.01$ & $\sigma_{\mathbf{w}} = 0.03$ 
 & $\sigma_{\mathbf{w}} = 0.01$ & $\sigma_{\mathbf{w}} = 0.03$ \\
\midrule
TV          & 26.74 / 0.846 & 25.62 / 0.803 & 29.16 / 0.902 & 26.17 / 0.820 & 29.36 / 0.904 & 26.05 / 0.815 \\
WCRR        & 27.25 / \underline{0.859} & 26.09 / \underline{0.819} & 30.26 / 0.920 & 26.88 / 0.838 & 30.38 / 0.922 & 26.75 / 0.833 \\
MFoE        & \textbf{27.36} / \textbf{0.860} & \textbf{26.18} / \underline{0.819} & \underline{30.43} / \underline{0.921} & \underline{27.14} / \underline{0.848} & \underline{30.65} / \underline{0.926} & \underline{27.04} / \underline{0.844} \\
Prox-DRUNet & \underline{27.34} / \underline{0.859} & \underline{26.17} / \textbf{0.820} & \textbf{30.55} / \textbf{0.924} & \textbf{27.24} / \textbf{0.852} & \textbf{30.78} / \textbf{0.928} & \textbf{27.15} / \textbf{0.848} \\
\bottomrule
\end{tabular}
\end{table*}

\subsection{Compressed-Sensing MRI}

We use fully sampled real-valued knee images of size $\left(320 \times 320\right)$ from a fastMRI dataset \cite{fastMRI2018} as ground truth, including scans with (PDFS) and without (PD) fat suppression. The corresponding measurements are obtained by a subsampling of the 2D Fourier transform ($k$-space). The Cartesian sampling mask is defined by two parameters: the acceleration factor $M_{\text{acc}}$; and the center fraction $M_{\text{cf}}$. Specifically, the central $\lfloor 320M_{\text{cf}} \rfloor$ columns of $k$-space (those corresponding to low frequencies) are fully retained, while the remaining columns are uniformly subsampled, which results in a total of $\lfloor 320 / M_{\text{acc}} \rfloor$ selected columns. Gaussian noise with standard deviation $\sigma_{\mathbf w} = 0.01$ is added to the real and the imaginary parts of the measurements. Each method is initialized with the zero-filled image, the hyperparameters are tuned on a validation set of 10 images, and the performance is reported on a testing set of 50 images, all normalized to the range $[0, 1]$. 

\begin{figure*}
\centering
\includegraphics[scale=0.9]{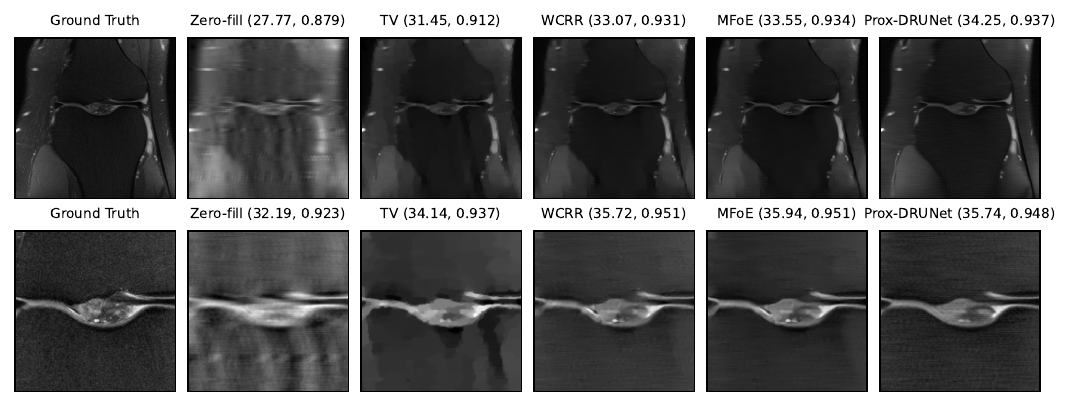}
\caption{Reconstruction for 8-fold subsampling and PDFS data. The reported metrics are (PSNR, SSIM).}
\label{fig:mri}
\end{figure*}

The performance is reported in Table \ref{tab:mri_results}. Again, the MFoE outperforms the WCRR in every instance. It also manages to outperform Prox-DRUNet on the $\left(M_{\text{acc}}=4, M_{\text{cd}}=0.08\right)$ instance with the PD data. Examples of reconstruction are provided in Figure~\ref{fig:mri}.

\begin{table}[t]
\centering
\caption{Quantitative results (PSNR / SSIM) for MRI reconstruction on the fastMRI dataset.}
\setlength{\tabcolsep}{3pt} 
\begin{tabular}{lcccc}
\toprule
 & \multicolumn{2}{c}{$(M_{\text{acc}}=4, M_{\text{cf}}=0.08)$} & \multicolumn{2}{c}{$(M_{\text{acc}}=8, M_{\text{cf}}=0.04)$} \\
\cmidrule(lr){2-3} \cmidrule(lr){4-5}
 & PD & PDFS & PD & PDFS \\
\midrule
Zero-fill   & 27.43 / 0.843   & 29.62 / 0.875   & 23.49 / 0.764   & 27.00 / 0.833 \\
TV          & 33.23 / 0.920 & 32.70 / 0.905 & 26.97 / 0.835 & 29.42 / 0.858 \\
WCRR        & 35.10 / \underline{0.950} & 34.18 / \underline{0.928} & 29.32 / 0.892 & 31.00 / \underline{0.888} \\
MFoE        & \textbf{35.40} / \textbf{0.952} & \underline{34.21} / \underline{0.928} & \underline{30.22} / \underline{0.903} & \underline{31.21} / \underline{0.888} \\
Prox-DRUNet & \underline{35.32} / \underline{0.950} & \textbf{34.53} / \textbf{0.931} & \textbf{30.97} / \textbf{0.904} & \textbf{31.66} / \textbf{0.894} \\
\bottomrule
\end{tabular}
\label{tab:mri_results}
\end{table}

\subsection{Computed Tomography}

We evaluate our method on the LoDoPaB-CT dataset \cite{LeuSch21lodopabct}, which contains ground-truth images derived from 2D slices of reconstructions in the LIDC/IDRI database \cite{armato2011lung}, each one of size $\left(362 \times 362\right)$. The measurement operators are defined by parallel-beam geometry and implemented using the ASTRA Toolbox \cite{astra, astra2}. All experiments are conducted with 256 simulated detectors. The resulting sinograms are corrupted with Gaussian noise of standard deviation $\sigma_{\mathbf w} = 0.1$.

We consider three acquisition settings with 60, 40, and 20 projection angles, respectively. Each method is initialized with the filtered-back-projected (FBP) image, and the hyperparameters are tuned on a validation set of 10 images. The performance is reported on a testing set of 50 images, all normalized to the range $[0, 1]$. 

\begin{figure*}
\centering
\includegraphics[scale=0.9]{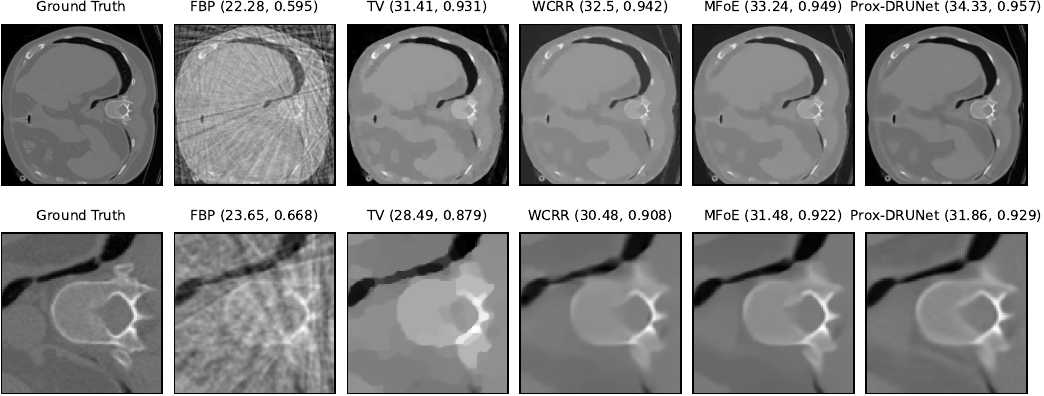}
\caption{Reconstruction for CT with 20 angles and 256 line detectors. The reported metrics are (PSNR, SSIM).}
\label{fig:ct}
\end{figure*}

The performance is presented in Table~\ref{tab:ct_results}. There too, MFoE outperforms WCRR in all tested configurations. Examples of reconstruction are provided in Figure~\ref{fig:ct}.

\begin{table}[t]
\centering 
\caption{Quantitative results (PSNR / SSIM) for CT reconstruction on the LoDoBaP dataset.}
\begin{tabular}{lccc}
\toprule
 & 60 angles & 40 angles & 20 angles\\ 
\hline
FBP         & 28.53 / 0.796   & 25.87 / 0.691   & 21.41 / 0.494 \\
TV          & 33.74 / 0.917 & 32.58 / 0.903 & 30.11 / 0.867 \\
WCRR        & 35.21 / 0.924 & 33.87 / 0.912 & 31.08 / 0.880 \\
MFoE        & \underline{35.40} / \underline{0.925} & \underline{34.21} / \underline{0.914} & \underline{31.48} / \underline{0.885} \\
Prox-DRUNet & \textbf{35.73} / \textbf{0.927} & \textbf{34.61} / \textbf{0.917} & \textbf{32.12} / \textbf{0.893} \\
\bottomrule
\end{tabular}
\label{tab:ct_results}
\end{table}

\subsection{Statistical Significance}

Across all our inverse problem experiments, we observe that WCRR outperforms TV, MFoE outperforms WCRR, and Prox-DRUNet usually outperforms MFoE. We assess the validity of this ranking through Wilcoxon signed-rank tests~\cite{wilcoxon1945individual} on the PSNR and SSIM scores across the entire test set, with a strict significance threshold of $p < 0.001$. In total, we perform 26 separate tests, covering the PSNR and SSIM metrics across all 13 experimental modalities (six deblurring, four CS-MRI, and three CT). 

Our testing framework confirms the robustness of this hierarchy. WCRR yields statistically significant improvements over TV in every single configuration. The transition from WCRR to MFoE is similarly distinct, with MFoE surpassing WCRR in all PSNR evaluations and achieving significant SSIM gains in 10 out of 13 modalities. Finally, Prox-DRUNet proves superior to MFoE in the majority of scenarios; statistical significance is established in 19 of the 26 comparisons, with the few exceptions occurring primarily in deblurring and CS-MRI tasks where both models achieved comparable high-fidelity reconstructions.

\subsection{Time Comparisons}

Table~\ref{tab:times} contains the average inference time per image, measured in seconds, for each inverse problem. More detailed results are provided in Tables~\ref{tab:deblurring_time}--\ref{tab:ct_time}. These results highlight that the performance gains of Prox-DRUNet come at a computational cost. In particular, the CT reconstructions with Prox-DRUNet require the repeated evaluation of the proximal operator of the data-fitting term—a step that is especially costly in the case of CT. This results in Prox-DRUNet being significantly slower than the other methods. The MFoE and WCRR methods are on average more than 13 times faster than Prox-DRUNet.

\begin{table}[t]
\centering 
\caption{Average duration [s] per image for all the reconstruction modalities.}
\begin{tabular}{lccc}
\toprule
  & Deblurring & CS-MRI & CT\\ 
\hline
TV     & 5.86 & 7.16 & 24.26 \\
WCRR   & 6.45 & 6.89 & 13.90 \\
MFoE   & 5.90 & 10.94 & 10.26 \\
Prox-DRUNet  & 39.82 & 62.52 & 267.42 \\
\bottomrule
\end{tabular}
\label{tab:times}
\end{table}

Although deep-learning-based methods such as Prox-DRUNet set a high standard in learned regularization, they do so at the cost of a large number of parameters and extensive training data. We believe that MFoE offers an interesting alternative with significantly fewer parameters, fewer training data, faster inference time, and with a reconstruction quality that is close to Prox-DRUNet. Together, these methods offer complementary strength and provide valuable options suited to a variety of application needs.

\section{Conclusion}

We have proposed the MFoE model. It is a principled extension of the classic FoE model and incorporates multivariate potential functions via Moreau envelopes of the $\ell_\infty$-norm. The image quality that it achieves noticeably narrows the performance gap with state-of-the-art learned regularizers while it offers a clear structured design that is easy to implement with competitive reconstruction times.

\section*{Acknowledgements}

The research leading to these results was supported by the European Research Council (ERC) under European Union’s Horizon 2020 (H2020), Grant Agreement - Project No. 101020573 FunLearn.

\bibliographystyle{IEEEtran}
\bibliography{references}

\appendix

\subsection{Proof of Theorem 1}\label{sec:proof}

We omit subscripts for clarity. By definition, we have that 

\begin{align}
    & \lefteqn{\hspace{-0.5em}\rho_{\tau\mu}^d(\mathbf{Qx}) = \inf_{\mathbf{z} \in \mathbb{R}^d} \left( \|\mathbf z\|_\infty + \frac{1}{2\tau\mu} \|\mathbf{z} - \mathbf{Qx}\|_2^2 \right)} \notag \\
    & \leq \|\mathbf Q\operatorname{prox}_{\mu \|\cdot\|_\infty}(\mathbf x)\|_\infty + \frac{1}{2\tau\mu}\|\mathbf Q\operatorname{prox}_{\mu \|\cdot\|_\infty}(\mathbf x) - \mathbf{Qx}\|_2^2 \notag \\
    & \leq \|\operatorname{prox}_{\mu \|\cdot\|_\infty}(\mathbf x)\|_\infty + \frac{1}{2\mu}\frac{\|\mathbf Q\|_2^2}{\tau}\|\operatorname{prox}_{\mu \|\cdot\|_\infty}(\mathbf x) - \mathbf{x}\|_2^2 \notag \\
    & = \|\mathbf x - \mu\operatorname{Proj}_{\mathcal{B}_{\ell_1}}(\mathbf x / \mu)\|_\infty + \frac{\mu}{2}\frac{\|\mathbf Q\|_2^2}{\tau}\|\operatorname{Proj}_{\mathcal{B}_{\ell_1}}(\mathbf x / \mu)\|_2^2. \label{eq:nonnegative}
\end{align}
It follows that 

\begin{align}\label{eq:diff_moreau}
    & \rho_\mu^d(\mathbf x) - \rho_{\tau\mu}^d(\mathbf{Qx}) \nonumber\\
    & \geq \frac{\mu}{2}\left(1 - \frac{\|\mathbf Q\|_2^2}{\tau}\right)\|\operatorname{Proj}_{\mathcal{B}_{\ell_1}}(\mathbf x / \mu)\|_2^2 \nonumber\\
    & \geq 0.
\end{align}
This lower bound is nonnegative for all $\mathbf{x}$. When $\mathbf{x} = \mathbf{0}$, the two Moreau envelopes in \eqref{eq:diff_moreau} vanish. This proves that $\psi(\mathbf x)$ is nonnegative with a unique zero-valued global minimum at the origin.

We now consider the gradient of $\psi(\mathbf{x})$ given by
\begin{equation}
    \nabla \psi(\mathbf{x}) = \mu \operatorname{Proj}_{\mathcal{B}_{\ell_1}}(\mathbf{x}/\mu) - \mu \mathbf{Q}^T \operatorname{Proj}_{\mathcal{B}_{\ell_1}}(\mathbf{Qx}/\tau \mu).
\end{equation}
Let $P = \mu \operatorname{Proj}_{\mathcal{B}_{\ell_1}}(\cdot / \mu)$ be the proximal operator of the indicator function of the set $\left\{ \mathbf{z} \in \mathbb{R}^d : \|\mathbf{z}\|_1 \leq \mu \right\}$.
It is therefore a firmly nonexpansive mapping \cite{rockafellar_prox}, so that
\begin{equation}
     \forall\, \mathbf{x}, \mathbf{y} \in \mathbb{R}^d: \quad \|P(\mathbf{x}) - P(\mathbf{y})\|_2^2 \leq \langle P(\mathbf{x}) - P(\mathbf{y}), \mathbf{x} - \mathbf{y} \rangle.
\end{equation}
We show that $\mathbf{Q}^T \operatorname{Proj}_{\mathcal{B}_{\ell_1}}(\mathbf Q \cdot /\tau \mu) = \mathbf{Q}^T P( \mathbf Q \cdot /\tau)$ is also firmly nonexpansive. This comes from the relations

\begin{align}
    & \lefteqn{\hspace{-2em}\left\|\mathbf{Q}^T P\!\left(\frac{\mathbf{Qx}}{\tau}\right) - \mathbf{Q}^T P\!\left(\frac{\mathbf{Qy}}{\tau}\right)\right\|_2^2} \notag\\
    & \leq \|\mathbf{Q}\|_2^2 
    \left\|P\!\left(\frac{\mathbf{Qx}}{\tau}\right)
    - P\!\left(\frac{\mathbf{Qy}}{\tau}\right)\right\|_2^2 \notag\\
    &\leq  \|\mathbf{Q}\|_2^2 \left\langle 
    P\!\left(\frac{\mathbf{Qx}}{\tau}\right) - P\!\left(\frac{\mathbf{Qy}}{\tau}\right),
    \frac{\mathbf{Qx}}{\tau} - \frac{\mathbf{Qy}}{\tau} \right\rangle \notag\\
    &= \frac{\|\mathbf{Q}\|_2^2}{\tau} \left\langle
    \mathbf{Q}^T P\!\left(\frac{\mathbf{Qx}}{\tau}\right)
    - \mathbf{Q}^T P\!\left(\frac{\mathbf{Qy}}{\tau}\right),
    \mathbf{x} - \mathbf{y} \right\rangle \nonumber\\
    &\leq \left\langle
    \mathbf{Q}^T P\!\left(\frac{\mathbf{Qx}}{\tau}\right)
    - \mathbf{Q}^T P\!\left(\frac{\mathbf{Qy}}{\tau}\right),
    \mathbf{x} - \mathbf{y} \right\rangle. \label{eq:firm-nonexp-qt} 
\end{align}
As shown in \cite{bauschke2011convex}, every firmly nonexpansive operator is half averaged, meaning there exists nonexpansive mappings $H_1$, $H_2$ such that $P = \frac{1}{2}\operatorname{Id} + \frac{1}{2}H_1$ and $\mathbf{Q}^T P(\mathbf Q \cdot /\tau \mu) = \frac{1}{2}\operatorname{Id} + \frac{1}{2}H_2$.
Applying this decomposition to $\nabla \psi$, we get that

\begin{align}
& \lefteqn{\hspace{-2em}\|\nabla \psi(\mathbf{x}) - \nabla \psi(\mathbf{y})\|_2} \notag \\
& \leq \tfrac{1}{2} \|H_1(\mathbf{x}) - H_1(\mathbf{y})\|_2 + \tfrac{1}{2} \|H_2(\mathbf{x}) - H_2(\mathbf{y})\|_2 \notag \\
& \leq \|\mathbf{x} - \mathbf{y}\|_2, 
\end{align}
which proves that $\nabla \psi$ is nonexpansive. $\square$

\subsection{Proof of Theorem 2}\label{sec:proof2}

Since the function $f$ is piecewise quadratic, it satisfies the Kurdyka-\L ojasiewicz property \cite{attouch_convergence_2013}. We rely on \cite[Theorem 3.7]{ipiano} which defines the conditions for the convergence of our algorithm. We define the sequence $(\mathbf z_k)_{k \in \mathbb N} = (\mathbf x_k, \mathbf x_{k-1})_{k \in \mathbb N}$ and the Lyapunov function $F(\mathbf z_k) = f(\mathbf x_k) + \delta \|\mathbf x_k - \mathbf x_{k-1}\|_2^2$. For $\{\mathbf z_k\}_{k \in \mathbb N}$ to converge to a critical point of $f$, three conditions must be met:

\begin{enumerate}
    \item there exists $a > 0$ such that, for any $k \in \mathbb N$,
    \begin{equation}
        F(\mathbf z_{k+1}) + a \|\mathbf x_k - \mathbf x_{k-1}\|^2_2 \leq F(\mathbf z_k);
    \end{equation}
    \item there exists $b > 0$ such that, for any $k \in \mathbb N$,
    \begin{equation}
        \|\mathbf \nabla F(\mathbf z_{k+1})\|_2 \leq \frac{b}{2}\left(\|\mathbf x_{k+1} - \mathbf x_k\|_2 + \|\mathbf x_{k} - \mathbf x_{k-1}\|_2\right);
    \end{equation}
    \item there exists a subsequence $\{\mathbf z_{k_j}\}_{j \in \mathbb N}$ such that
    \begin{equation}
        \mathbf z_{k_j} \to \mathbf z^* \quad \text{and} \quad F(\mathbf z_{k_j}) \to F(\mathbf z^*).
    \end{equation}
\end{enumerate}

For the first two conditions, we focus solely on the case where the inertial step is accepted,
\begin{equation}
    \mathbf x_{k+1} = \mathbf x_k - \alpha \mathbf \nabla f(\mathbf x_k) + \beta_k(\mathbf x_k - \mathbf x_{k-1}),
\end{equation}
and note that the standard gradient step has already been proven to satisfy these conditions in \cite{ipiano}. \\

\noindent \textbf{Condition 1.} By the definition of our algorithm, we have that
\begin{align}
    F(\mathbf z_{k+1}) &- F(\mathbf z_k) \leq - \alpha \left(1 - \tfrac{L\alpha}{2}\right)\|\nabla f(\mathbf x_k)\|^2 \nonumber \\
    &+ \delta \|\mathbf x_{k+1} - \mathbf x_k \|^2 - \delta \|\mathbf x_k - \mathbf x_{k-1}\|^2.
\end{align}
The Expansion of the norm term yields that
\begin{align}
    \|\mathbf x_{k+1} - \mathbf x_k\|^2 &= \alpha^2 \|\nabla f(\mathbf x_k)\|^2 + \beta_k^2 \|\mathbf x_k - \mathbf x_{k-1}\|^2 \nonumber \\
    &\quad - 2\alpha\beta_k \langle \nabla f(\mathbf x_k), \mathbf x_k - \mathbf x_{k-1} \rangle,
\end{align}
which leads to
\begin{align}\label{eq:proof2_eq1}
    F&(\mathbf z_{k+1}) - F(\mathbf z_k) \leq - \alpha \left(1 - \tfrac{L\alpha}{2} - \delta \alpha\right)\|\nabla f(\mathbf x_k)\|^2 \nonumber \\
    &+ \delta(\beta_k^2 - 1) \|\mathbf x_k - \mathbf x_{k-1}\|^2 \nonumber \\
    &- 2\delta\alpha\beta_k \langle \nabla f(\mathbf x_k), \mathbf x_k - \mathbf x_{k-1} \rangle.
\end{align}
Applying Young's inequality with $\lambda > 0$, we have that
\begin{multline}\label{eq:proof2_eq2}
    - 2\alpha\beta_k \langle \nabla f(\mathbf x_k), \mathbf x_k - \mathbf x_{k-1} \rangle \\
    \leq 2 \alpha |\beta_k| \left(\frac{1}{2\lambda} \|\nabla f(\mathbf x_k)\|^2 + \frac{\lambda}{2} \|\mathbf x_k - \mathbf x_{k-1}\|^2\right).
\end{multline}
The combination of \eqref{eq:proof2_eq1} with \eqref{eq:proof2_eq2} leads to
\begin{align}\label{eq:proof2_eq3}
    F(\mathbf z_{k+1}) &- F(\mathbf z_k) \leq - \alpha \left(1 - \tfrac{L\alpha}{2} - \delta \alpha - \tfrac{\delta |\beta_k|}{\lambda}\right)\|\nabla f(\mathbf x_k)\|^2 \nonumber \\
    &+ \delta(\beta_k^2 - 1 + \alpha\lambda|\beta_k|) \|\mathbf x_k - \mathbf x_{k-1}\|^2.
\end{align}
We select $\lambda = \delta |\beta_k| / (1 - L\alpha/2 - \delta \alpha)$ to eliminate the gradient term. Since $\alpha < 2/L$, we can ensure that $\lambda > 0$ by choosing $\delta \in (0, \tfrac{1}{\alpha} - \tfrac{L}{2})$. This simplifies Inequality~\eqref{eq:proof2_eq3} to
\begin{equation}
    F(\mathbf z_{k+1}) - F(\mathbf z_k) \leq \delta\left(\beta_k^2 - 1 + \tfrac{\delta\alpha\beta_k^2}{1 - L\alpha/2 - \delta\alpha}\right)\|\mathbf x_k - \mathbf x_{k-1}\|^2.
\end{equation}
To satisfy Condition 1, the coefficient on the right-hand side must be negative. Given $\delta > 0$, this requires that
\begin{equation}
    \beta_k^2 \left(1 + \tfrac{\alpha \delta}{1 - L\alpha/2 - \delta\alpha}\right) = \beta_k^2 \tfrac{1 - L\alpha/2}{1 - L\alpha/2 - \delta\alpha}< 1.
\end{equation}
Since the numerator and denominator are positive, this is equivalent to
\begin{equation}
    \beta_k^2 < \frac{1 - L\alpha/2 - \delta\alpha}{1 - L\alpha/2} \leq 1.
\end{equation}
Because $\delta$ can be chosen arbitrarily small, the upper bound approaches $1$. Thus, the condition holds if $\beta_k \in (-1, 1)$, which is guaranteed by Algorithm~\ref{alg:hbr}. \\

\noindent \textbf{Condition 2.} We observe that
\begin{align}
    \|&\nabla F(\mathbf z_{k+1})\| \nonumber \\
    &\leq \|\nabla f(\mathbf x_{k+1}) + 2 \delta(\mathbf x_{k+1} - \mathbf x_k)\| + \|2 \delta(\mathbf x_{k+1} - \mathbf x_k)\| \nonumber \\
    &\leq \|\nabla f(\mathbf x_{k+1})\| + 8 \delta^2 \|\mathbf x_{k+1} - \mathbf x_k\| \nonumber \\
    &\leq \|\nabla f(\mathbf x_{k+1}) - \nabla f(\mathbf x_k)\| + \|\nabla f(\mathbf x_k)\| \nonumber \\
    &\quad + 8 \delta^2 \|\mathbf x_{k+1} - \mathbf x_k\| \nonumber \\
    &\leq L \|\mathbf x_{k+1} - \mathbf x_k\| + \tfrac{1}{\alpha} \|\mathbf x_{k+1} - \mathbf x_k\| + \tfrac{\beta_k}{\alpha}\|\mathbf x_k - \mathbf x_{k-1}\| \nonumber \\
    &\quad + 8\delta^2 \|\mathbf x_{k+1} - \mathbf x_k\| \nonumber \\
    &= \left(L + \tfrac{1}{\alpha} + 8 \delta^2\right)\|\mathbf x_{k+1} - \mathbf x_k\| + \tfrac{\beta_k}{\alpha}\|\mathbf x_k - \mathbf x_{k-1}\| \nonumber \\
    &\leq \left(L + \tfrac{1}{\alpha} + 8 \delta^2 + \tfrac{\beta_k}{\alpha}\right) \left(\|\mathbf x_{k+1} - \mathbf x_k\| + \|\mathbf x_k - \mathbf x_{k-1}\|\right).
\end{align}
The condition is satisfied by setting $b/2$ equal to the strictly positive constant term $\left(L + \tfrac{1}{\alpha} + 8 \delta^2 + \tfrac{\beta_k}{\alpha}\right)$. \\

\noindent \textbf{Condition 3.} As established in the first condition, the value of $F(\mathbf z_k)$ is non-increasing. Since the sequence $\{\mathbf z_k\}_{k \in \mathbb N}$ is bounded, by the Bolzano-Weierstrass theorem, there exists a subsequence $\mathbf z_{n_k}$ converging to $\mathbf z^*$. Finally, the continuity of $F$ implies that $F(\mathbf z_{n_k}) \to F(\mathbf z^*)$. $\square$

\subsection{Parameterization of $\{\mu_k\}_{k=1}^K$}\label{sec:noise_level}

We recall that the $\mu_k\colon \mathbb R \to \mathbb R$, which depend on the noise level $\sigma$, are the Moreau envelope parameters in Equation~\eqref{eq:diff_moreau}. They are concatenated in the function $\boldsymbol\mu = (\mu_k)_{k=1}^K$, specified as 

\begin{equation}
    \boldsymbol\mu(\sigma) = \frac{1}{100}\left(\sigma \cdot \mathbf{1}_K + \frac{1}{20}\mathbf f(\bar{\sigma})\right)_+ + 10^{-9}\cdot\mathbf{1}_K,
\end{equation}
where $\mathbf f\colon \mathbb R \to \mathbb R^K$ is a three-layer neural network with a hidden dimension $K$. The leading term in this formula scales linearly with the noise level, while the neural network $\mathbf{f} \colon \mathbb{R} \to \mathbb{R}^K$ provides a nonlinear correction. The function $(\cdot)_+ \coloneqq \max(0, \cdot)$ and the addition of a small term are there to ensure the positivity of the components of $\boldsymbol\mu$. The term $\bar{\sigma}$ represents the noise level normalized to the interval $[-2, 2]$. This normalization matches the standard input distribution expected by the neural network at initialization. The factor $1/100$ calibrates the parameter for image intensities in $[0, 1]$. The factor $1/20$ essentially forces the network to start as a small perturbation around a linear law ($\boldsymbol\mu \propto \sigma$). This ensures that the training starts from a physically plausible state (linear scaling) rather than from a random one.

\subsection{Implementation of WCRR-Free}\label{sec:wcrr_free}

To implement WCRR-free, we closely follow the approach of \cite{23M1565243}, parameterizing the potential function $\psi$ via its derivative $\varphi$ as
\begin{equation}
\varphi = \mu_{+} \varphi_{+} - \mu_{-} \varphi_{-},
\end{equation}
where $\mu_+, \mu_- > 0$ are learnable scalar weights and $\varphi_+$, $\varphi_-$ are trainable, nondecreasing, nonexpansive linear splines. The only deviation from \cite{23M1565243} is that we allow $\mu_-$ to be learnable, whereas it was fixed to $\mu_- = 1$ in the original implementation. We assign to $\mu_-$ the same learning rate as $\mu_+$.

\subsection{Implementation of MFoE-$\ell_2$}\label{sec:mfoe_l2}

Let $\eta^d_\mu \colon \mathbb{R}^d \to \mathbb{R}_+$ denote the Moreau envelope of the $\ell_2$-norm. Using \eqref{eq:moreau}–\eqref{eq:prox_proj}, it can be expressed as

\begin{equation}
    \eta^d_\mu(\mathbf x) = \|\mathbf x - \mu\operatorname{Proj}_{\mathcal{B}_{\ell_2}}(\mathbf x / \mu)\|_{2} + \frac{\mu}{2}\|\operatorname{Proj}_{\mathcal{B}_{\ell_2}}(\mathbf x / \mu)\|_2^2,
\end{equation}
where

\begin{equation}
    \mathcal{B}_{\ell_2} = \left\{ \mathbf{z} \in \mathbb{R}^d : \|\mathbf{z}\|_2 \leq 1 \right\}
\end{equation}
denotes the unit $\ell_2$-ball in $\mathbb{R}^d$. We build the potentials as

\begin{equation}\label{eq:potential_l2}
    \psi(\mathbf x) = \mu\eta_{\mu}^d(\mathbf x) - \mu\eta_{\tau\mu}^d(\mathbf Q \mathbf x).
\end{equation}
To ensure the same guarantees as in Theorem~\ref{thm:single_global_min}, it is sufficient to impose that $\|\mathbf Q\|_2 \leq 1$ and $\tau > 1$. With the exact same mathematical manipulations as \eqref{eq:nonnegative}, we can show that

\begin{align}
    & \eta_\mu^d(\mathbf x) - \eta_{\tau\mu}^d(\mathbf{Qx}) \nonumber\\
    & \geq \frac{\mu}{2}\left(1 - \frac{\|\mathbf Q\|_2^2}{\tau}\right)\|\operatorname{Proj}_{\mathcal{B}_{\ell_2}}(\mathbf x / \mu)\|_2^2 \nonumber \\
    & \geq 0.
\end{align}
This allows us to draw the same conclusions as in Section~\ref{sec:proof}. The gradient of $\psi(\mathbf{x})$ given by
\begin{equation}
    \nabla \psi(\mathbf{x}) = \mu \operatorname{Proj}_{\mathcal{B}_{\ell_2}}(\mathbf{x}/\mu) - \mu \mathbf{Q}^T \operatorname{Proj}_{\mathcal{B}_{\ell_2}}(\mathbf{Qx}/\tau \mu)
\end{equation}
can also be proven to be nonexpansive with the exact same arguments as in Section~\ref{sec:proof}.

\subsection{Additional Experimental Results}

A comparison of various configurations of the MFoE is provided in Table~\ref{tab:denoising-results3} for a denoising task. The best performance occurs at $(K=15, d=4)$, which is the setting used in all other experiments of this paper.

\begin{table}
\centering
\caption{Denoising performance on the BSD68 test set.}
\setlength{\tabcolsep}{5pt}
\begin{tabular}{lccc}
\toprule
& $\sigma=15/255$ & $\sigma=25/255$ & $\sigma=50/255$ \\
\hline
MFoE ($K=60$, $d=1$) & 31.19 & 28.68 & 25.76 \\
MFoE ($K=30$, $d=2$) & 31.31 & 28.82 & 25.91 \\
MFoE ($K=20$, $d=3$) & 31.31 & 28.83 & 25.90 \\
MFoE ($K=15$, $d=4$) & \textbf{31.32} & \textbf{28.84} & \textbf{25.92} \\
MFoE ($K=12$, $d=5$) & 31.30 & 28.82 & 25.90 \\
MFoE ($K=10$, $d=6$) & 31.29 & 28.81 & 25.90 \\
MFoE ($K=6$, $d=10$) & 31.26 & 28.78 & 25.87 \\
MFoE ($K=5$, $d=12$) & 31.26 & 28.80 & 25.89 \\
MFoE ($K=4$, $d=15$) & 31.23 & 28.77 & 25.84 \\
MFoE ($K=3$, $d=20$) & 31.19 & 28.73 & 25.81 \\
MFoE ($K=2$, $d=30$) & 31.16 & 28.71 & 25.81 \\
MFoE ($K=1$, $d=60$) & 31.01 & 28.59 & 25.70 \\
\bottomrule
\end{tabular}
\label{tab:denoising-results3}
\end{table}

\subsection{Time Comparisons}

A comparison of the computation times of various image reconstruction frameworks across different modalities is given in Tables~\ref{tab:deblurring_time}, \ref{tab:mri_time}, and \ref{tab:ct_time}.

\begin{table*}
\centering
\caption{Average reconstruction time [s] per image in deblurring.}
\begin{tabular}{lcccccccc}
\toprule
  & \multicolumn{2}{c}{\includegraphics[scale=1.0]{figures/kernel1.pdf}} & \multicolumn{2}{c}{\includegraphics[scale=1.0]{figures/kernel2.pdf}} & \multicolumn{2}{c}{\includegraphics[scale=1.0]{figures/kernel3.pdf}} \\ & $\sigma_{\mathbf w} = 0.01$ & $\sigma_{\mathbf w} = 0.03$ & $\sigma_{\mathbf w} = 0.01$ & 
  $\sigma_{\mathbf w} = 0.03$ & $\sigma_{\mathbf w} = 0.01$ & 
  $\sigma_{\mathbf w} = 0.03$\\ 
\hline
TV & 9.70 $\pm$ 2.44 & 5.39 $\pm$ 1.26 & 5.93 $\pm$ 1.26 & 3.47 $\pm$ 0.70 & 6.32 $\pm$ 1.18 & 4.33 $\pm$ 0.77 \\
WCRR & 4.75 $\pm$ 1.70 & 4.95 $\pm$ 1.41 & 7.03 $\pm$ 1.60 & 7.35 $\pm$ 1.86 & 7.54 $\pm$ 1.47 & 7.07 $\pm$ 1.88 \\
MFoE & 4.09 $\pm$ 0.97 & 4.07 $\pm$ 1.03 & 6.92 $\pm$ 2.07 & 6.14 $\pm$ 1.45 & 6.48 $\pm$ 2.01 & 7.70 $\pm$ 1.91\\
Prox-DRUNet  & 21.87 $\pm$ 11.47 & 20.33 $\pm$ 11.76 & 37.42 $\pm$ 16.86 & 45.17 $\pm$ 15.19 & 39.16 $\pm$ 15.57 & 74.95 $\pm$ 22.51 \\
\bottomrule
\end{tabular}
\label{tab:deblurring_time}

\vspace{1em}

\caption{Average reconstruction time [s] per image in CS-MRI.}
\begin{tabular}{lcccc}
\toprule
  & \multicolumn{2}{c}{$\left(M_{\text{acc}}=4, M_{\text{cf}}=0.08\right)$} & \multicolumn{2}{c}{$\left(M_{\text{acc}}=8, M_{\text{cf}}=0.04\right)$} \\  & PD & PDFS & PD & PDFS\\ 
\hline
TV           & 4.89 $\pm$ 0.98 & 6.83 $\pm$ 1.28 & 9.93 $\pm$ 1.60 & 7.00 $\pm$ 1.60 \\
WCRR         & 5.72 $\pm$ 1.71 & 4.46 $\pm$ 1.34 & 10.91 $\pm$ 1.90 & 6.48 $\pm$ 1.83 \\
MFoE & 7.84 $\pm$ 1.93 & 4.90 $\pm$ 1.14 & 18.74 $\pm$ 5.10 & 12.28 $\pm$ 4.26\\
Prox-DRUNet  & 73.32 $\pm$ 20.55 & 41.06 $\pm$ 15.69 & 67.78 $\pm$ 16.08 & 67.90 $\pm$ 14.06 \\
\bottomrule
\end{tabular}
\label{tab:mri_time}

\vspace{1em}

\caption{Average reconstruction time [s] per image in CT.}
\begin{tabular}{lccc}
\toprule
  & 60 angles & 40 angles & 20 angles\\ 
\hline
TV     & 24.12 $\pm$ 2.59 & 25.02 $\pm$ 2.93 & 23.63 $\pm$ 1.45 \\
WCRR   & 13.96 $\pm$ 1.82 & 14.16 $\pm$ 1.31 & 13.58 $\pm$ 2.18 \\
MFoE & 9.09 $\pm$ 3.03 & 9.09 $\pm$ 2.35 & 12.61 $\pm$ 3.42\\
Prox-DRUNet  & 266.98 $\pm$ 209.95 & 308.04 $\pm$ 124.73 & 368.30 $\pm$ 126.36 \\
\bottomrule
\end{tabular}
\label{tab:ct_time}
\end{table*}

\end{document}